**ONSET OF GIANT PLANET MIGRATION BEFORE 4480 MILLION YEARS AGO**

Stephen J. Mojzsis[1,2], Ramon Brasser[3], Nigel M. Kelly[1], Oleg Abramov[4] and Stephanie C. Werner[5]

[1]Department of Geological Sciences, University of Colorado, UCB 399, 2200 Colorado Avenue, Boulder, Colorado 80309-0399, USA

[2]Institute for Geological and Geochemical Research, Research Centre for Astronomy and Earth Sciences, Hungarian Academy of Sciences, 45 Budaörsi Street, H-1112 Budapest, Hungary.

[3]Earth Life Science Institute, Tokyo Institute of Technology, Ookayama, Meguro-ku, Tokyo 152-8550, Japan.

[4]Planetary Science Institute, 1700 East Fort Lowell Road, Suite 106, Tucson, AZ 85719, USA.

[5]Centre for Earth Evolution and Dynamics, University of Oslo, Postbox 1028 Blindern, N-0315 Oslo, Norway

Editorial correspondence:

Stephen J. Mojzsis

University of Colorado

Department of Geological Sciences

2200 Colorado Ave., Boulder CO 80309-0299 USA

email: mojzsis@colorado.edu; telephone: 1-303-492-5014; fax: 1-303-492-2606

**Abstract**

Immediately after their formation, the terrestrial planets experienced intense impact bombardment by comets, left-over planetesimals from primary accretion, and asteroids. This temporal interval in solar system evolution, termed *late accretion*, thermally and chemically modified solid planetary surfaces and may have impeded life's emergence on the Hadean (pre-3.85 Ga) Earth. The sources and tempo of late accretion are, however, vague. Here, we present a timeline that relates variably retentive radiometric ages from asteroidal meteorites, to new dynamical models of late accretion that invokes giant planet migration. Reconciliation of the geochronological data with dynamical models shows that giant planet migration immediately leads to an intense ~30 Myr influx of comets to the entire solar system. The absence of whole-sale crustal reset ages after ~4450 Ma for the most resilient chronometers from Earth, Moon, Mars, Vesta and various meteorite parent bodies confines the onset of giant planet migration to no later than ca. 4480 Ma. Waning impacts from planetesimals, asteroids (and a minor cometary component) continue to strike the inner planets through a protracted monotonic decline in impactor flux; this is in agreement with predictions from crater chronology. Amended global 3-D thermal analytical bombardment models derived from our new impact mass-production functions show that persistent niches for prebiotic chemistry on the early Hadean Earth could endure late accretion for at least the last 4400 Myr. (222 words)

1. INTRODUCTION

The nature of early bombardment to the inner solar system is a long-standing problem in planetary science. In a general sense, *late accretion* refers to the long period of bombardment wherein leftover material composed of comets, planetesimals and asteroids struck the planets after their formation. Although a matter of debate, a signal of this bombardment has been traced by the abundance of highly siderophile elements (HSE, which include the platinum group elements: Os, Ir, Ru, Rh, Pt, and Pd) as pollutants in a planet's mantle (Walker 2009), which are in turn used to infer an exogenous mass augmentation after separation of its core and mantle.

There exist two major paradigms that describe the time-dependent flux of late accretion. The first argues for a burst of relatively short duration in the impact bombardment intensity hailed as a "late heavy bombardment" (LHB), which affected the whole inner solar system (Morbidelli et al. 2012, 2018; cf. Ryder 2002) at around 3950 Ma, or perhaps earlier. The LHB is largely implied by interpretations of $^{40-39}$Ar chronology from lunar samples (e.g. Fernandes et al. 2013) and has been promulgated by several popular dynamical models involving late-stage giant planet migration (e.g. Gomes et al. 2005). Alternatively, the second concept proposes an overall monotonic decline in the bombardment rate, an interpretation most broadly consistent with crater counting studies (Hartman 1970; Neukum et al. 1975; Werner & Ivanov 2015). To transcend the inconsistency between the two different models for late accretion – which also has a strong bearing on the Hadean Earth's proclivity to host a nascent biosphere (e.g. Maher & Stevenson 1988) – requires a comprehensive review of geochronological data for the inner solar system, and a new investigation of planetary dynamics for the outer solar system.

2. LATE ACCRETION: MONOTONIC DECLINE OR IMPACT CATACLYSM?

Direct records of the first ~100 Myr of solar system evolution that could shed light on the inception of late accretion appear to be absent from the crusts of the Moon, Earth and Mars. On Earth, this deficit has been ascribed to endogenous crust-mantle interactions (e.g. Hopkins et al. 2008, 2010), or less popularly, to the destructive thermal and mechanical effects of impact bombardments that re-homogenized and as a result reset the silicate reservoirs (crust and mantle). Either way, Earth's terrestrial silicate reservoirs equilibrated rather late, at approximately 4480 Ma, well after primary accretion had ceased as attested by analysis of terrestrial samples using time-sensitive tracers of lithosphere production and destruction such as Pb-, Xe-, and Nd-isotopes (e.g. Albarède &

Juteau 1984; Allégre et al. 1995; Zhang 2002). For the Moon, lunar zircons (zirconium orthosilicate, or $Zr(SiO_4)$) show that its crust probably did not completely solidify until at least ca. 4417 Ma (Nemchin et al. 2009). Remote sensing studies indicate Venus' crust may be no older than about 1000 Ma (McKinnon et al. 1997; cf. Rolf et al. 2018), and the oldest features on Mercury may be comparable in age to those of the Moon and Mars (Marchi et al. 2013). The firmest *minimum* date for the perpetuation of the crusts of Earth and Mars comes from preserved ages for the oldest terrestrial zircons at ca. 4400 Ma (Valley et al. 2014) and the oldest comparable martian zircons, which have been dated at 4428-4470 Ma (Bouvier et al. 2018). The mere existence of these minerals demonstrates that the crusts of the sampled terrestrial worlds did not experience wholesale melting since that time. Thus, available geological data from the terrestrial planets that bear on the temporal flux profile of late accretion are confined to the last ca. 4450 Myr.

Earth's Moon shows an unusual history: Impact-modified lunar samples returned by the Apollo and Luna missions from the lunar near side yield $^{40-39}$Ar, U-Pb and Rb-Sr ages that seem to cluster around 3950 Ma (Turner et al. 1973; Tera et al. 1974); this is far younger than the suggested solidification age of the lunar crust cited above (Nemchin et al. 2009). Such a grouping of ages has long been used to argue that a cataclysmic bombardment affected the inner solar system as a rapid rise, and ensuing demise, in impactor flux (Ryder 1990). Intriguingly, neither lunar, mercurian nor martian crater chronologies show evidence for such an LHB event (Werner 2014), and it has been suggested that the 3950 Ma ages may instead be from a biased sampling compromised from widespread contamination of debris associated with the Moon's Imbrium basin (e.g. Shearer & Borg 2006). Statistical analyses of crater distributions instead point to a *monotonic* decline in impactor flux from at least 4400 Ma **(Figure 1)**. Hence, it is equivocal whether there was an LHB at all, or that if there was, it occurred with as yet no clear temporal constraints (e.g. Hartmann 2003).

Beyond the Earth, Moon, and Mars, documented highly-retentive U-Pb and Pb-Pb ages in eucrite meteorites attributed to asteroid 4Vesta (dubbed HEDs, for the howardite-eucrite-diogenite group of meteorites and collectively termed *vestoids*) record a solid crust at 4563 Ma (Ireland & Wlotzka 1992), and subsequent thermal modifications until ca. 4450 Ma (Hopkins et al. 2015). Practically no such ages younger than ca. 4450 Ma exist, however, for any known meteorite class including the vestoids. Younger ages for meteorite classes across the board are instead largely confined to less retentive $^{40-39}$Ar geochronology datasets (Bogard 1995), which generally display age continua from ~4450 Ma extending to about 3000 Ma, with occasional resetting events up to now (e.g Fernandes et al. 2013).

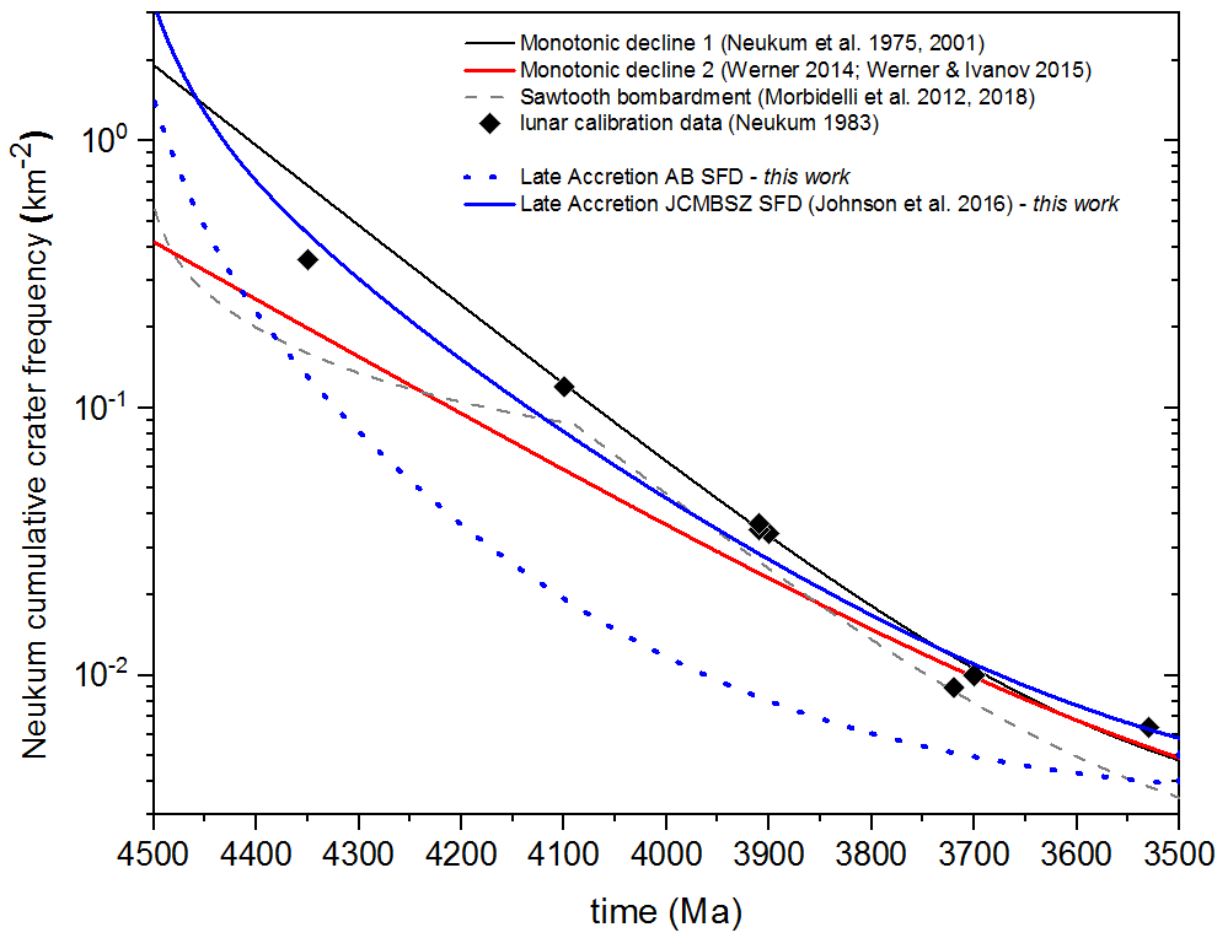

**Figure 1**. Lunar impact fluxes from various sources. The differential number of lunar craters with diameter > 1 km N(1) as a function of time and per unit surface for several scenarios discussed in the text. AB SFD = "asteroid belt size-frequency distribution". JCMBSZ SFD = "Johnson-Collins-Minton-Bowling-Simonson-Zuber size-frequency distribution".

Despite these numerous discrepancies, the lunar crater record continues its historical role as the standard reference for the nature of bombardment experienced by all solar system bodies (Öpik 1960; Shoemaker et al. 1962). Recent re-analyses, however, of purported $^{40\text{-}39}$Ar age clusters at ca. 3950 Ma from lunar rocks are inconclusive as to whether they were caused by an LHB, or are the result of a monotonic decline (Boehnke & Harrison 2016). As such, the problem of late accretion reduces to deciding for *which* period can the lunar crater record be used to interpret *what* primordial impact history connected with the ongoing accretion process after planet formation. Furthermore, *how* translatable is the lunar record to other solid surfaces of the solar system? Progress in decipherment of this record would greatly enhance our understanding of the formation and evolution of the solar system as a whole, and the thermal histories of the crusts of the terrestrial planets; the latter includes the Earth's capability to host a persistent biosphere from the earliest times (Mojzsis et al. 1996; Abramov & Mojzsis 2009; Abramov et al. 2013).

### 3. THE ROLE OF THE GIANT PLANETS

Notwithstanding debate over the existence of the LHB, a suite of dynamical models has been constructed to explain how a discrete and temporary surge in impactors could have transpired at 3950 Ma, or earlier (e.g. Gomes et al., 2005; Bottke et al. 2012; Marchi et al. 2012; Morbidelli et al. 2012, 2018; Nesvorný et al. 2018). The common theme to these models is that they rely on a dynamical instability in the outer solar system. Indeed, support for orbital migration of the giant planets (Jupiter, Saturn, Uranus and Neptune) in the first few hundred million years comes from several independent lines of evidence stemming from the orbital properties of these planets themselves, and analysis of small body populations beyond Neptune (Nesvorný, 2018). Clement et al. (2019) provide a recent example of such dynamical simulations that invoke a very early instability. Without temporal limits placed by geochronological data, however, caution is warranted in accepting such models at face value even if an early instability can naturally reproduce the small mass of Mars (there are other models that can do this; e.g. Brasser et al. 2017).

The giant planets are thought to have radially migrated because of their gravitational interactions with a massive population of planetesimals (Fernandez & Ip 1984). The migration of Neptune is chronicled in the orbital distribution of Pluto and the Kuiper Belt objects (Malhotra 1993, 1995). Dynamical models also show that giant planet migration requires planet-planet scattering (Thommes et al. 1999; Brasser et al. 2009; Levison et al. 2011; Nesvorný et al. 2012). Giant planet migration scenarios all share the prediction that the inner solar system should experience a short but intense surge of cratering by material that originated mostly in the outer solar system (comets) sometime in the first 500 Myr (Bottke et al. 2007).

Two points must be emphasized: (1) The specific trigger for giant planet migration – and ensuing bombardment – is unknown; and, (2) its timing is inexplicit, and could occur after a delay of a few million years to upwards of several hundred million years (Levison et al. 2013). Although debated, clear evidence for the role of outer solar system materials (i.e. comets) in this migration-induced bombardment is lacking on the Moon (Kring & Cohen, 2002; Strom et al. 2005 cf. Greenwood et al. 2011). Be that as it may, all dynamical models of terrestrial planet formation envisage a steady decline in bombardment rate once migration ceases (Neukum et al. 2001; Brasser et al. 2016). This expectation for late accretion is consistent with the crater chronologies cited above, but is apparently at odds with an LHB-like "spike" at any time after about 4400 Ma (Hartmann et al. 2000).

For instance, if giant planet migration commenced sometime after 4400 Ma to give rise to a discrete LHB-type event, it should be apparent as a distinct cluster of younger ages shared by all inner solar system bodies – including the Moon – in radiogenic systems relatively sensitive to thermal re-setting, such as in $^{40-39}$Ar and Rb-Sr (e.g. Bogard 1995; Fernandes et al. 2013). If the migration commenced before 4400 Ma, endogenous crustal processes on geologically active worlds such as Earth and Mars should have erased such evidence, but not necessarily for a small, airless, old and cold body like asteroid 4Vesta (Zhou et al. 2013).

The classical LHB case for a cataclysmic spike affecting the whole solar system after about 4100 Ma (Ryder 2002) is further undermined by numerous documented pre-4100 Ma (and older) lunar, vestoid, martian and terrestrial ages. These ages are surmised to reflect partial or complete resetting of high closure-temperature mineral geochronological systems (e.g. Norman & Nemchin 2014; Hopkins & Mojzsis 2015; Kelly et al. 2018). Closure temperature (hereafter, $T_c$) is defined as the temperature of a system at the time given by its apparent age (Dodson 1973; Reiners et al. 2005). Taken together, it is evident that at least from the lunar record, no firm conclusions can be drawn about the proposed mechanics of the early bombardment process, nor of the precise timing of giant planet migration that caused it. A new approach is warranted, and for this we turn to the asteroids.

## 4. EFFECTS OF LATE ACCRETION TO THE ASTEROIDS

The diverse members of the asteroid belt (DeMeo & Carry 2014) completed their formation before the inner planets and are thus witnesses to the solar system's earliest post-primary accretion history. We use compiled chronological data for asteroidal meteorites and compare different radiogenic systems with different $T_c$ in different minerals and whole-rock samples. This approach circumvents the debate between the monotonic decline vs. LHB cataclysm scenarios cited above because asteroidal meteorite ages are not anchored to the lunar record, and there is no dispute that they preserve a history that pre-dates the Moon's own formation. Robust chronometers like U-Pb in zircon from meteorites such as the brecciated eucrites – a subset of vestoids that have undergone substantial reworking through impacts – document thermal re-set ages (Hopkins et al. 2015) for a system with substantially higher closure temperature ($T_c$ ~1000 °C; [e.g. Reiners et al. 2015]) than, for example, $^{40-39}$Ar (250-450 °C). Minerals that host radiogenic systems with intermediate $T_c$, such as Rb-Sr and U-Pb in phosphate (e.g. basic Ca-phosphate, or apatite), are expected to record ages of intermediate bombardment intensity, thus bridging the gap between the highly-retentive U-Pb in zircon and the low-retentivity of $^{40-39}$Ar **(Figure 2)**. Accordingly, $T_c$ provides a clear theoretical basis for understanding meteorite mineral and whole-rock ages as cooling ages arising from the interplay between the kinetics of diffusion (or annealing) and accumulation rates in cooling radio-isotopic systems from processes such as the thermal consequences of impacts.

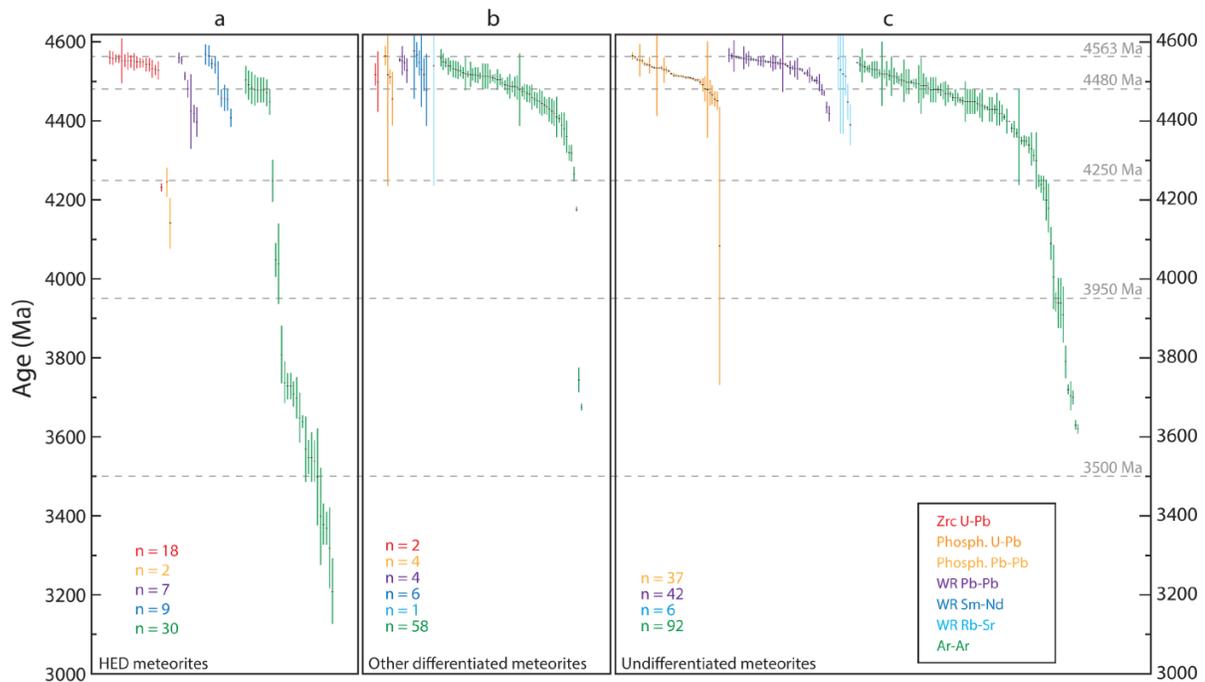

**Figure 2.** Distributions of impact-reset ages of meteorites of different meteorite class (HED, other differentiated meteorites, undifferentiated meteorites). Each datum represents the centers and widths corresponding to the absolute age and 2σ errors of each dated sample. The different colored groupings of data correspond to different geochronological data with different retentivities (legend).

Our thesis is simple: High $T_c$ highly-retentive radiogenic systems will chiefly record older ages as opposed to the less-retentive radiogenic systems. The most retentive systems such as U-Pb in zircon (when available) will only ever record their own formation in a melt, or (rarely) the most energetic events either caused by higher impact velocities or by impacts from larger bodies. The prediction is that U-Pb in zircon registers ages of crust formation, or reset ages when the impact intensity is so high that the crust is molten. The $^{40-39}$Ar system will only retain ages when the impact flux was low(er) and the solar system had dynamically relaxed.

An LHB-style surge in the impact spike from giant planet migration at any time in solar system history should re-set ages for the asteroid belt, and appear as a cluster of ages in a variety of radiogenic systems, not just $^{40-39}$Ar.

To understand this further and explain the data in **Figure 2**, we combine our chronological compendium from meteorites in **Supporting Information (Table S2)** with dynamical models to better constrain when the giant planets could have commenced their migration. This task requires that we model the temporal impact bombardment history onto the terrestrial planets by accounting for the major asteroids, leftover planetesimals of terrestrial planet formation, and comets arriving from the outer solar system. The outcomes of these models for late accretion, linked as they are to the geochronological constraints imposed herein, are then used to revise previous analyses of the thermal consequences of impact bombardments to the crusts of the planetary bodies of the inner solar system. Special focus is placed on the habitable potential of the Hadean Earth.

## 5. DYNAMICS OF LATE ACCRETION

The bombardment history of the terrestrial planets consists of three components: (i) dynamical evacuation of the (E-belt) asteroids, (ii) a contribution from leftover planetesimal material from terrestrial planet formation, and (iii) an influx of comets is summarized in **Figure 3**. In the following sections, we discuss the dynamical setup that determines the amount of late accretion onto the Moon and the terrestrial planets from E-belt asteroids and leftover planetesimals. The contribution from comets is discussed separately below owing to the fact that the dynamical model is much more involved.

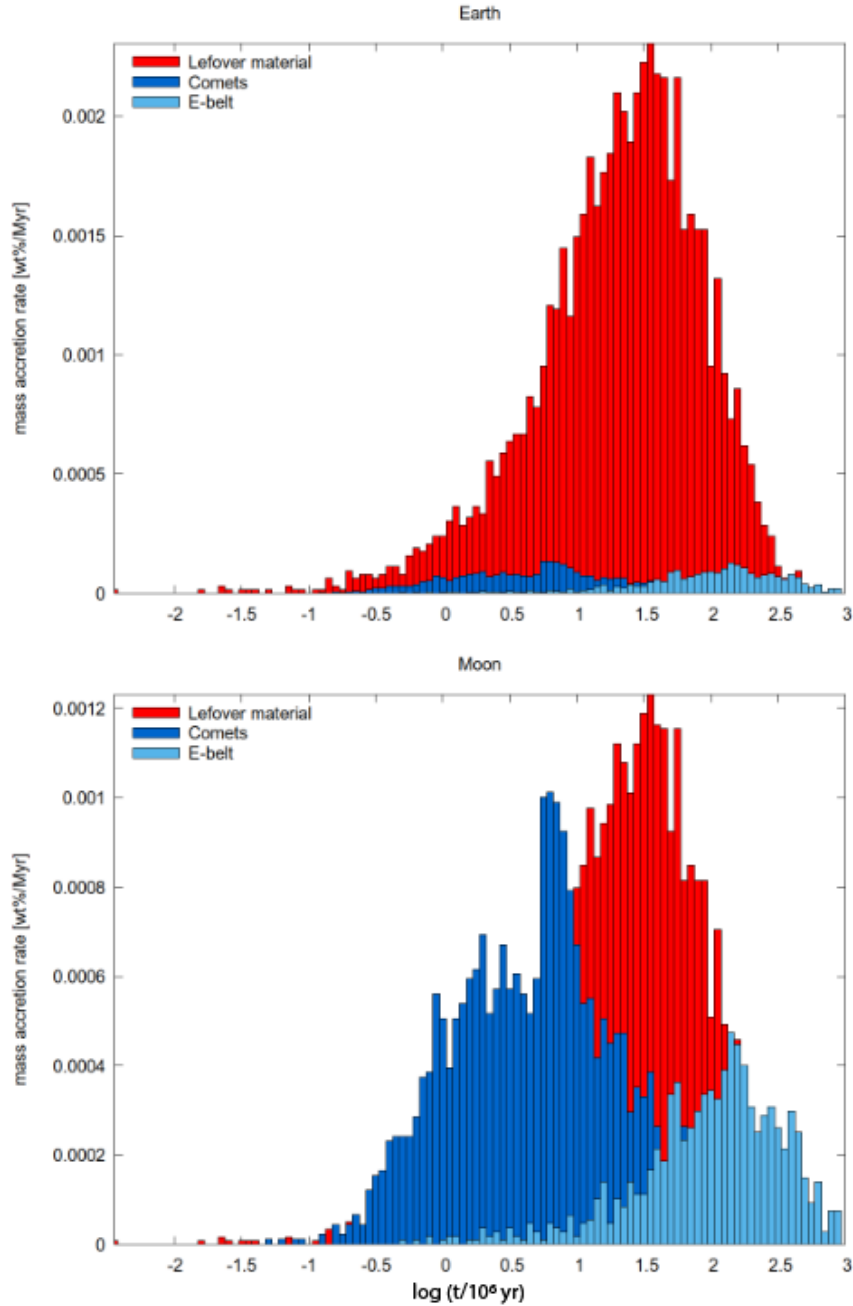

**Figure 3.** Impact rates onto the Earth and Moon from comets, leftover planetesimals and the E-belt expressed as mass accretion rate in weight percent/Myr on target (Earth, Moon) vs. time expressed in log/Myr.

### 5.1 The E-belt

The E-belt is a now-empty implied extension of the asteroid belt all the way to Mars [56]. We compute the contribution to late accretion from the E-belt by running 32 dynamical N-body simulations of the evolution of this reservoir in the presence of the gravitational influence of the Sun and the eight planets on their current orbits. We also added asteroids 4Vesta and 1Ceres as massive bodies. Each simulation also has 512 massless asteroids. We employ the following initial conditions [36]: the asteroids have a uniform distribution in semi-major axes *a* from 1.7–2.1 AU and a main-asteroid-belt-like Gaussian distribution in eccentricity *e* and inclination *i*, with mean values of 0.15 and 8.5°, respectively, and standard deviations of 0.07 and 7°, respectively. The other three angles (longitude of the ascending node, argument of perihelion and mean anomaly) are chosen uniformly random between 0° and 360°. The differences between each simulation are the different initial conditions of the planetesimals, even though these are statistically identical. We also ensured that initially none of the asteroids were on Mars-crossing orbits.

The simulations were run for 1 Gyr of model time with SWIFT RMVS 3 (Levison & Duncan 1994) with a time step of 3.65 d. Planetesimals were removed once they were further than 100 AU from the Sun (whether bound or unbound) or when they collided with a planet or with the Sun. The impact probabilities were independently obtained from counting direct impacts onto the planets and from employing a post-processing procedure (Wetherill 1967) that computes the average impact probability. The latter is the only source to compute the impact probability with the Moon. We have previously tested this computational method against counting direct impacts with the planets and we find that the two results match those in Bottke et al. (2012). The impact probabilities and average impact velocities for the terrestrial planets, 4Vesta and 1Ceres are listed in **Table 1**. Analysis of the dynamics of late accretion to the other terrestrial planets is forthcoming in future work.

We caution that the average impact velocity increases with time, and that the values listed here are averaged over 1000 Myr. We also list the expected amount of mass striking the surfaces of Earth, Moon and Vesta from Monte Carlo impact experiments (Brasser et al. 2016) and their 2σ uncertainties (see below). The uncertainties obtained from the assumed shallow size-frequency distribution are much larger than that in the impact probabilities. We further computed that the impact probability of objects from the main asteroid belt with the Moon is at least a factor of 30 lower than that of the E-belt, consistent with earlier estimates (Morbidelli et al. 2010). Since the two reservoirs are thought to initially have a similar mass (Bottke et al. 2005) the contribution from the asteroid belt (E-belt) is thus insignificant in the first 1000 Myr and henceforth neglected in our analysis.

| Planetary body | $P_{imp}$ [%] | $V_{imp}$ [km s$^{-1}$] | Expected mass striking the surface [wt.%] |
|---|---|---|---|
| Mercury | 0.9 | 41 | $0.0067^{+0.017}_{-0.0063}$ |
| Venus | 4.9 | 26 | $0.0027^{+0.0077}_{-0.0024}$ |
| Earth | 5.4 | 21 | $0.0024^{+0.0060}_{-0.0020}$ |
| Moon | 0.25 | 20 | $0.0090^{+0.013}_{-0.0087}$ |
| Mars | 2.2 | 11 | $0.0080^{+0.024}_{-0.0072}$ |
| 4Vesta | 0.016 | 9 | 0.15 |
| 1Ceres | 0.0022 | 11 | 0.0056 |

**Table 1.** Impact probabilities, average impact velocities and mass accreted from E-belt material with the terrestrial planets, Moon, 4Vesta and 1Ceres.

We caution that the average impact velocity increases with time, and that the values listed here are averaged over 1000 Myr. We also list the expected amount of mass striking the surfaces of Earth, Moon and Vesta from Monte Carlo impact experiments (Brasser et al. 2016) and their 2σ uncertainties (see below). The uncertainties obtained from the assumed shallow size-frequency distribution are much larger than that in the impact probabilities. We further computed that the impact probability of objects from the main asteroid belt with the Moon is at least a factor of 30 lower than that of the E-belt, consistent with earlier estimates (Morbidelli et al. 2010). Since the two reservoirs are thought to initially have a similar mass (Bottke et al. 2005) the contribution from the asteroid belt (E-belt) is thus insignificant in the first 1000 Myr and henceforth neglected in our analysis.

### 5.2  Leftover planetesimals

We have previously constrained the mass in leftover planetesimals from terrestrial planet formation (Brasser et al. 2016) based on the HSE abundance in the lunar mantle (Walker 2009). We then used that information to show that the purported "Late Veneer" (an early, albeit significant component of late accretion) to the Earth and Mars consisted mostly of single, large impactors of dominantly enstatite or ordinary chondrite composition (Brasser et al. 2018; Genda et al. 2017; Woo et al. 2018, 2019). Owing to the fact that our analysis of leftover material requires that the Moon was present, we start our dynamical simulations at the time of the Moon-forming event, assigned here at 4510 Ma [Barboni et al. 2017; cf. Connelly et al. 2016). We also possess a database of terrestrial planet formation simulations in the framework of the Grand Tack model (Walsh et al. 2011). Snapshots of the planetesimal population were taken 60 Myr after the start of the simulations, which we approximately assign at 4500 Ma. We split the total number of planetesimals into 32 cases of 1074 massless planetesimals and added all the major planets, plus 4Vesta and 1Ceres, on their current orbits. Simulations were run with SWIFT RMVS3 for 500 Myr with a time step of 3.65 d with the same removal conditions described above for the E-belt simulations. Results for the Earth, Moon and Vesta are provided in **Table 2**. Uncertainties are 2σ.

| Planet | $P_{imp}$ [%] | $V_{imp}$ [km s$^{-1}$] | Expected mass striking the surface [wt.%] |
|---|---|---|---|
| Mercury | 8.2 | 31 | $0.52^{+1.49}_{-0.51}$ |
| Venus | 32 | 23 | $0.13^{+0.20}_{-0.11}$ |
| Earth | 24 | 19 | $0.13^{+0.17}_{-0.11}$ |
| Moon | 0.8 | 17 | 0.032 (fixed) |
| Mars | 2.9 | 13 | $0.12^{+0.23}_{-0.11}$ |
| 4Vesta | 0.0021 | 11 | 0.19 |
| 1Ceres | 0.0027 | 11 | 0.071 |

**Table 2.** Average impact probabilities and impact velocities of leftover planetesimals with the terrestrial planets, Moon, 4Vesta and 1Ceres.

No direct record exists of how much mass was in these leftover planetesimals. Lunar HSE abundances show that the Moon had to have accreted a further 0.025 wt.% after its formation while it still had a magma ocean (Day et al. 2015). Estimates for the duration of a lunar magma ocean vary (Boyet et al. 2007; Borg et al. 2011; Elkins-Tanton et al. 2011; Kamata et al. 2015) but if we assume 80 Myr and we take into account the decline in the number of planetesimals from the dynamical simulations, then the Moon accreted an additional 0.00625 wt.% after it formed a crust, for a total of 0.032 wt.%. Combined with the average impact probability with the Moon and employing a size-frequency distribution akin to that of the main asteroid belt suggests the mass in leftover planetesimals at 4500 Ma is approximately $4.7^{+4.4}_{-3.7} \times 10^{-3}$ Earth masses. To study the effect of the size-frequency distribution on the mass accreted by the planets we rely on Monte Carlo impact experiments.

We recently combined N-body and Monte Carlo simulations to establish the nature of late accretion onto the terrestrial planets, and the likely total mass accreted by the Earth and Mars when calibrated to lunar HSE abundances (Brasser & Mojzsis 2017). Due to the assumed shallow size-frequency distribution of the impactors, the ratio of the total mass accreted by the Earth versus that of the Moon is generally much higher than the ratio of their gravitational cross sections (Bottke et al. 2010). The accreted mass is dominated by stochastic accretion from a few large impactors (Tremaine & Dones 1993). We presupposed the bodies in the simulations had minimum diameters of 1 km, a maximum diameter of 2000 km and a bulk density of 2500 kg m$^{-3}$ for all objects regardless of their size. Two cases are tested: we either presume the same shallow size-frequency distribution as that of the main asteroid belt, or with a changeover to a steeper cumulative slope of 4.5 at a projectile diameter of 50 km (Johnson et al. 2016). For further details of the method we refer the reader to (Brasser et al., 2016). The expected values of material accreted by the Earth and Mars are much lower than their actual values of 0.7 wt% and 0.8 wt% respectively (Day et al. 2016); these outcomes are discussed extensively in Hartmann et al. (2000). We did not include Vesta and Ceres in the Monte Carlo code because they are not expected to be struck by large bodies, so the listed values are estimates obtained from the dynamical simulations. When considering only impactors with a maximum diameter of 2000 km, the accreted amounts in **Table 2** for Mercury, Venus, Earth and Mars should be divided by 2 (as well as the uncertainties).

*5.3 Comets*

The cometary impact flux onto the terrestrial planets is caused by the late migration of the giant planets. By *late* we mean well after the gas of the protosolar nebula has dissipated; we assume the gas disk is dispersed by ca. 4563 Ma based on the youngest chondrule ages (e.g. Morris et al. 2015; Bollard et al. 2017). Late giant planet migration is caused by the scattering of distant planetesimals beyond Neptune and by mutual scattering among the giant planets (Tsiganis et al. 2005).

We computed the cometary flux onto the terrestrial planets in three sequential steps. *Step 1*: Find the best initial conditions for the giant planets and the planetesimal disk that ultimately has the highest probability of reproducing the current architecture of the terrestrial planets. *Step 2*: Compute the fraction of comets from the trans-Neptunian disk that venture closer than 1.7

AU from the Sun as a proxy for objects that reach the terrestrial planets. *Step 3*: Compute the impact probability of these comets with the terrestrial planets. *Step 4*: Apply Monte Carlo impact simulations to the comets to compute the amount of material that strikes the Earth and the Moon.

These steps are now explained in detail.

Step 1: Initial conditions. The evolution of the giant planets during late migration is chaotic (Tsiganis et al. 2005), and the probability of the planets ending up near their current orbits is low: ~5% (Brasser & Lee 2015; Nesvorný 2015a). The aim is to increase that probability by varying the initial conditions and decide which combination of three input parameters best reproduces the current configuration of the giant planets. Here we follow the procedure outlined in Wong et al. (2019).

The initial conditions employed here are identical to a loose 5-planet configuration (Nesvorný 2011) because this configuration best reproduces the current orbital structure of the Kuiper belt (Nesvorný 2015b). To start, we assume there are five giant planets in the quadruple resonance 3:2, 3:2, 2:1, 3:2. The initial conditions for this configuration were provided by D. Nesvorný (*personal communication*), but we allowed the initial semi-major axis of Jupiter to vary (first free parameter); the initial semi-major axes of the other planets were computed from their resonant locations. Variations in the initial conditions within each set were obtained by giving a uniformly random deviation of $10^{-6}$ AU to the position vector of each planetesimal.

The planetesimal disk consisted of 3000 objects. The surface density of the planetesimal disk scales with heliocentric distance as $\Sigma \propto r^{-1}$. The total disk mass is the second free parameter. The inner edge of the disk is located 1 AU outward from the outermost ice giant; the outer edge of the disk is our third free parameter.

We run many sets of 128 simulations each wherein the three free parameters are permutated. We make use of the symplectic integrator SyMBA (Duncan et al. 1998). Planets and planetesimals were removed once they were further than 1000 AU from the Sun (whether bound or unbound) or when they collided with a planet or ventured closer than 0.5 AU to the Sun. We evolve every simulation to 500 Myr or until fewer than four planets remain, whichever occurs first. The migration evolution and final semi-major axes and eccentricities for the giant planets are depicted in **Figure 4**. We show cases for a planetesimal disc of 18 (red) and 19 Earth masses (blue). The black dots and error bars denote the current semi-major axes and eccentricities and the extent of their secular variation. The arrows indicate the direction and amount of migration. We conclude that the combination of parameters that yields the highest probability of reproducing the current solar system according to the criteria of Nesvorný (2015b) has Jupiter initially at 5.6 AU, a disc mass of 18 Earth masses ($M_⊕$) and an outer edge at 27 AU. These parameters are used as input for Step 2.

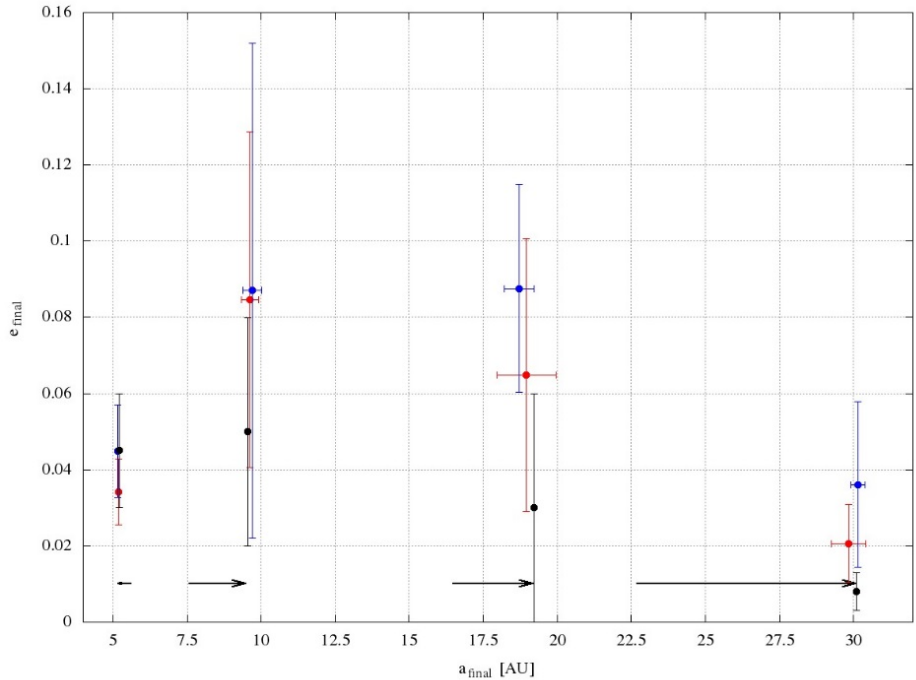

**Figure 4** Final semi-major axes and eccentricities of giant planets after migration. Arrows indicate amount and direction.

Step 2. Compute the fraction of comets that enter the inner solar system. In order to calculate the impact probability of comets onto the Moon and the terrestrial planets we use a piecewise approach and split the work across two sets of simulations.

We took the best initial conditions determined in Step 1 and ran one set of 512 simulations with the same initial configuration of the planets, disk mass and disc outer edge, but now comets were removed from the simulation when they ventured closer than 1.7 AU from the Sun. The vectors and the time of crossing 1.7 AU are stored for the next step.

We find that 22% of all comets from the trans-Neptunian disk crossed the 1.7 AU barrier, or the equivalent of ca. 4 $M_⊕$. The outcome of this experiment is used in step 3.

Step 3. Calculate the impact probability of the comets with the terrestrial planets. All the comets that crossed the 1.7 AU barrier were subsequently integrated for 10 Myr in simulations that included all eight planets on their current orbits, as well as Vesta and Ceres. This duration is about two orders of magnitude longer than the average dynamical lifetime of a Jupiter-family comet (Brasser & Morbidelli 2013). The comets were introduced into the simulation with the position and velocity when they crossed the barrier from Step 2. All runs were performed with SWIFT RMVS3 with the same time step and removal criteria as for the leftover planetesimals.

The comets are dynamically controlled by Jupiter and consequently they only ever spend little time in the inner solar system (Levison & Duncan 1997; Di Sisto et al. 2009). We therefore do not record any physical impacts onto the terrestrial planets, and resort to an averaging procedure (Wetherill 1967) that computes the impact probability and therefore the total accreted mass. **Table 3** shows the cometary impact probabilities with the Earth, the Moon and Vesta when taking into account the 22% probability of the comets venturing into the inner solar system from Step 2 herein. We also list the average impact velocity and the expected amount of mass in comets that strikes the surface of each of these bodies.

| Planet | $P_{imp}$ [x $10^{-6}$] | $V_{imp}$ [km s$^{-1}$] | Expected mass striking the surface [wt.%] |
|---|---|---|---|
| Mercury | 0.13 | 34 | 0.0041 |
| Venus | 1.8 | 26 | 0.0041 |
| Earth | 3.1 | 22 | $0.0055^{+0.00087}_{-0.00057}$ |
| Moon | 0.16 | 20 | $0.024^{+0.0041}_{-0.0035}$ |
| Mars | 2.4 | 13 | $0.043^{+0.00048}_{-0.00035}$ |
| 4Vesta | 0.0093 | 11 | 0.39 |
| 1Ceres | 0.025 | 11 | 0.29 |

**Table 3.** Impact probabilities and average impact velocities of cometary material with the terrestrial planets, Moon, 4Vesta and 1Ceres. Monte Carlo simulations were only run for Earth, Moon and Mars.

Step 4: Monte Carlo impact experiments. We apply the same Monte Carlo experiments to the comets to compute the amount of material that impacts the Earth and the Moon. The simplest size-frequency distribution that matches the Kuiper belt is used, with a change in slope at a diameter of ca. 60 km and cumulative slope indices of 4.8 and 1.9 respectively at the high and low ends (Fraser & Kavelaars 2009). Since the distribution exhibits a knee at diameters of 60 km most of the mass in the belt resides in objects of approximately this size and thus we expect that the ratio of accreted masses between the planets is approximately equal to their gravitational cross sections. We assumed a bulk density of 1400 kg m$^{-3}$ (~ 2× computed average comet density; Peale 1989). Outcomes of these analyses are shown in **Figures 5** and **6**.

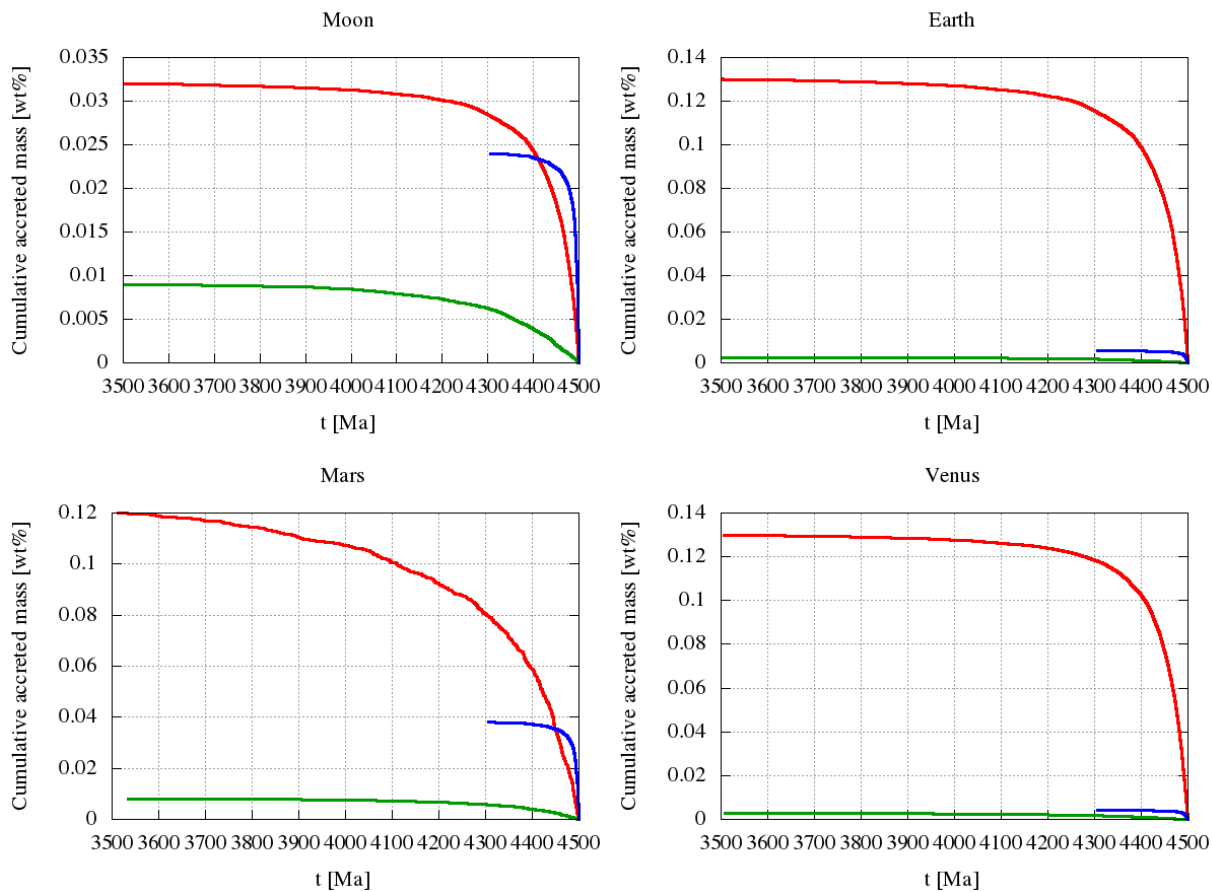

**Figure 5.** Accretion profiles for the Moon, Earth, Mars and Venus assuming the giant planet instability occurred at 4.5 Ga. This plot assumes that all impactors have the same mass. Red is for leftover planetesimals, blue is for comets, green for the E-belt.

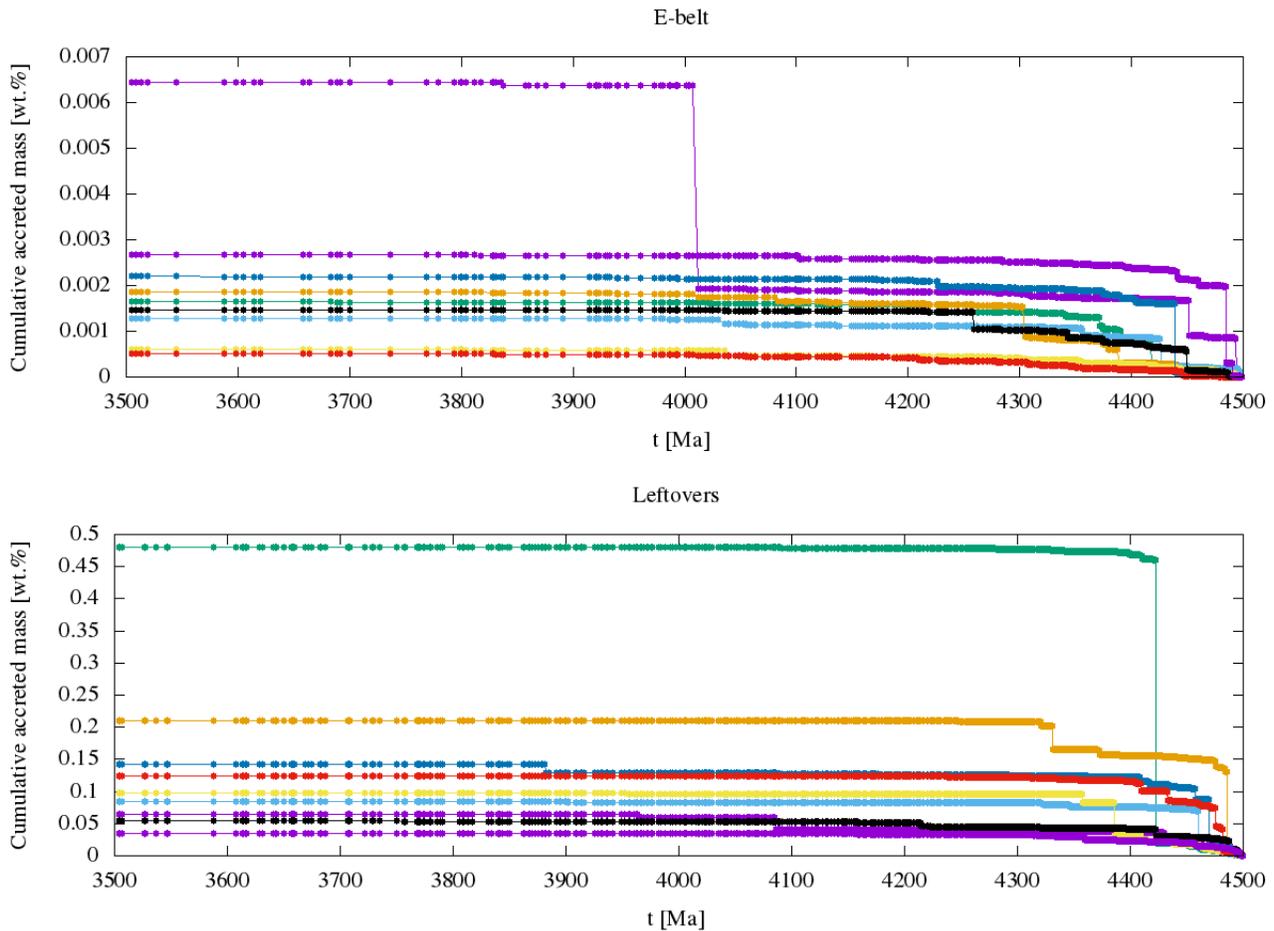

**Figure 6.** Late accretion onto Earth from the E-belt (top) and leftover material (bottom) taking the size-frequency distribution of the impactors into account.

**6. Global thermal models of late accretion to the Hadean Earth.**

The revised global thermal cratering model in **Figure 7** consists of (i) a stochastic cratering model which populates a cuboid representing the Hadean Earth's surface with craters using the mass-production functions derived from the results of our new dynamical models; (ii) analytical expressions that calculate a temperature field for each crater (Abramov & Mojzsis 2009; Abramov et al. 2013); (iii) a three-dimensional thermal model of the terrestrial lithosphere, where craters are allowed to cool by conduction and radiation; and, (iv) instantaneous plus cumulative melt production as well as defined thermal volumes of the crust as a function of time (Kieffer & Simonds 1980; Pierazzo & Melosh 2000). The model output reports the amount of melting in the lithosphere, both instantaneous and cumulative, as well as thermal volumes in the top 4 km of the lithosphere (surface to depth) at any point in time.

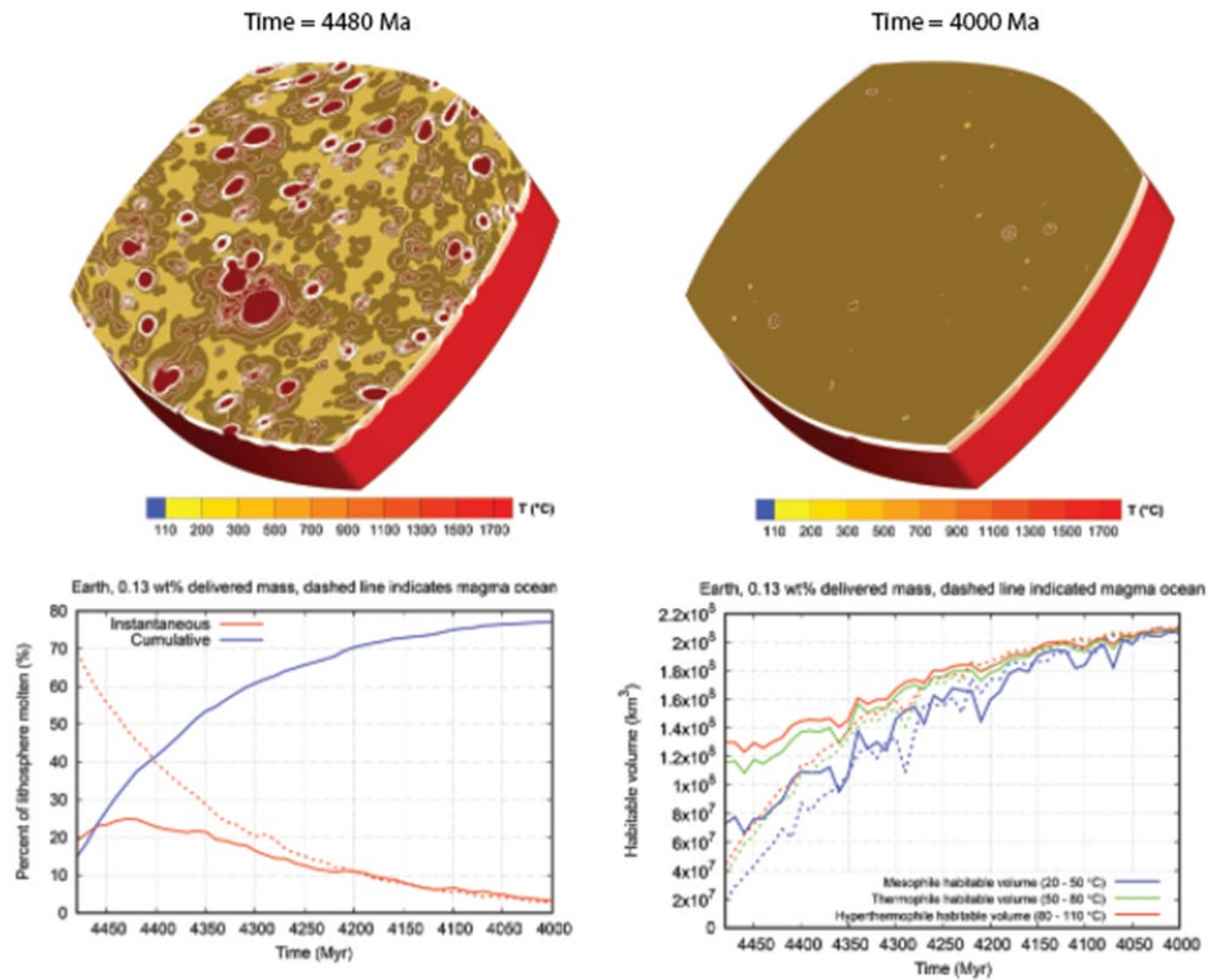

**Figure 7**. Three-dimensional thermal models representing the entire volume of Hadean Earth's lithosphere as a full wraparound cuboid at various times during late accretion beginning at 4480 Myr **(a)** and concluding at 4000 Myr ago **(b)**. Only impactors larger than 1 km in diameter are included in this representation; our baseline model was 0.13 wt.% of delivered mass to Earth over the course of late accretion. The upper surface shows temperatures at a depth of 4 km. Dark areas indicate crater imprints. The simulation extends to a depth of 150 km on the vertical scale. **(c)** Immediate thermal effects of impacts on the lithosphere. Percent of the Earth's lithosphere to experience melting at any given time during late accretion. The solid line represents the baseline model and the dashed line represents the model with an initial magma ocean condition. Lithospheric thickness has no significant effect on these results. **(d)** Thermal states of crustal volumes in hydrothermal environments. Solid lines represent the baseline model and the dashed lines represents the model with an initial magma ocean condition. Mesophile (20–50 °C, blue), thermophile (50–80 °C, green) and hyperthermophile (80–110 °C, red) thermal tolerance volumes are reported for ~500 Myr of bombardment and resultant active hydrothermal environments.

*6.1 Thermal consequences of late accretion to the Hadean Earth*

We simulate impact bombardment that Earth experienced between 4500 Myr and 4000 Myr based on the late accretion simulations presented herein. Mean surface temperature and geothermal gradient were assumed as 20 °C and 70 °C km$^{-1}$, and a basaltic composition was assigned for the initial lithosphere. Total delivered mass was 7.76 × 10$^{21}$ kg, equivalent to 0.13 % of the Earth's mass, for the baseline model, although 0.065 wt% delivered mass was also tested. The size-frequency of this late accretion impacts was approximated using the size-frequency distribution of the main asteroid belt. The velocity distribution of the impactors, and the decline of impact flux with time, was derived from dynamical modeling as described in this work. Impactor density of these rocky bodies was assumed to be 3000 kg m$^{-3}$, and impact angle of each impactor was stochastically generated from a Gaussian distribution centered at 45 degrees. Some of our simulations began with a magma ocean scenario. We deem this scenario as a likely consequence of the largest post-Moon formation impactors in late accretion implicated, for example, in the Earth's postulated Late Veneer (e.g. Genda et al. 2017). Based on previous work by us (Abramov & Mojzsis 2009) and others (Chyba 1990; Marchi et al. 2014), we estimate that a minimum total delivered mass of 9.31×10$^{21}$ kg of accreted material is required to melt the terrestrial lithosphere at any particular point in time. Assuming asteroid 1Ceres' density, this corresponds to an object of approximately 1200 km in diameter (about the size of Pluto's moon, Charon) striking the Earth to melt the crust. This condition is also satisfied by a multitude of smaller objects. At a 13% impact probability with the Earth, the total mass in leftover planetesimals that is required to account for melting the terrestrial lithosphere is 0.011 M$_\oplus$ at 4500 Ma; thish is well within acceptable parameter space (Brasser et al. 2016).

*6.2 Thermal effects of late accretion to asteroid 4Vesta*

Our dynamical simulations indicate that Vesta is struck by 0.39 wt% of cometary material (**Table 3**) which translates to 10$^{18}$ kg; the equivalent diameter of an object of that mass is about 110 km. The typical mass of the largest object to impact Vesta for the size-frequency distribution used here is ~1/3 of the total mass (Tremaine & Dones 1993). This suggests the diameter of the largest comet to have struck Vesta is about 76 km, or about 1/8 of the diameter of Vesta itself. If Vesta was struck by an object twice as large, we expect that such an impact could have mechanically disrupted the asteroid. Using a typical impact velocity of 11 km, an impact with an object of diameter 76 km would raise the temperature of about $2.5 \times 10^5$ km$^3$ of Vesta's crust (or about 0.35% of Vesta's total volume) above 1100 °C (see Figure 8 in Hopkins et al. 2015). Our estimate of Vesta's crustal resetting is a lower limit because twice as much more mass will strike its crust in smaller projectiles. If we naively assume that the heat spreads evenly across the asteroid's surface, this amount of volume would yield a layer of 286 m deep available to thermally reset any radiogenic ages in that volume. Therefore, if the cometary bombardment occurred in an LHB-like scenario, we would expect that the brecciated vestoid meteorites (eucrites) should show substantial *late* U-Pb reset ages in zircon around 4100 or 3950 Ma. They do not.

## 7. RESULTS AND DISCUSSION

Our compilation of asteroidal meteorite age data do not show a common LHB-like spike pattern near 3950 Ma, and we therefore rule out a specific solar system-wide bombardment at that time caused by giant planet migration. Anticipated LHB patterns do not exist at 4100-4200 Ma, nor are they present back to ca. 4400 Ma. Data instead show that the highly-retentive radiogenic U-Pb and Pb-Pb ages older than 4530 Ma - variably preserved in ordinary chondrites, vestoids, and other meteorite groups in **Figure 2** - are attributable to parent body formation ages (Ireland & Wlotzka 1992; Bogard 1995). Crustal processes inherent to these small bodies which could reset ages after 4530 Ma are precluded owing to the demise of the principle short-lived radioactive heat source to the early solar system $^{26}$Al (e.g. Lichtenberg 2016), and the dearth of other radioactive heat sources with sufficient power output (e.g. $^{60}$Fe, $^{244}$Pu). Hence, our preferred interpretation of ages substantially younger than 4530 Ma for any asteroidal meteorite is that they document impact-induced thermal resets past a particular $T_c$ of a radiogenic system from a protracted decline of late accretion.

Intriguingly, a number of meteorite classes display patterns of U-Pb and Pb-Pb in zircon and apatite age resets that terminate at about 4450 Ma (**Figure 2**). The whole-rock Pb-Pb and Sm-Nd systems, with intermediate $T_c$ of 600 °C (below that of U-Pb in zircon, but above $^{40-39}$Ar) terminate later, at around 4400 Ma. We find that $^{40-39}$Ar ages tend to pick up near 4480 Ma, roughly where U-Pb, Pb-Pb and Sm-Nd ages leave off, with little overlap between the groups. Argon data also show a continuum in ages extending from about 4480-3000 Ma, and younger. We view this behavior of sequential termination of ages in minerals with decreasing $T_c$ as the expected result of a long monotonic decline of leftover debris of planet formation, and the products of occasional inter-asteroid collisions (Bottke et al. 2005). The result is in agreement with crater counting statistics for the Moon, and generally comports with the meta-analysis of compiled lunar $^{40-39}$Ar data by Boehnke & Harrison (2016). It is decidedly at odds, however, with any bombardment model that calls for a cataclysmic LHB-style spike (or "spikes") in impact activity at any time in solar system history since about 4450 Ma.

It may also be the case that the relative rarity of resolved $^{40-39}$Ar ages older than ca. 4450 Ma for the asteroidal meteorites is a result of the "Stonewall Effect" (Hartmann et al. 2000; Hartmann 2003), in which a history of earlier thermal events becomes masked. In the Stonewall scenario, the last impacts energetic enough to re-set ages are recorded and earlier incidents tend to become so severely overprinted that they become effectively obliterated. The near-absence of reset ages younger than 4450 Ma in, for example, robust U-Pb and Pb-Pb chronologies for asteroidal meteorites also agrees with trends of declining bombardment intensity in line with our new dynamical models. Collectively, meteorite age distributions viewed through the lens of dynamical studies presented herein show that giant planet migration commenced before 4480 Ma, since the ensuing cometary bombardment records no cluster of ages at any time after.

*7.1        Constraining the timing of the onset of giant planet migration*

We expect that a signature for giant planet migration should be evident in traces of a brief but intense cometary input to the moons and planets as proposed by Gomes et al. (2005) and shown in **Figure 3**. For example, lunar hydrogen isotope (D/H) values suggest that the Moon accreted some cometary material after it formed (Greenwood et al. 2011, 2018). Xenon isotopes on both the Earth and Moon hint that these bodies acquired some part of their noble gas inventories from comets (Marty et al. 2017; Bekaert et al 2017); it has also been postulated that some Xe in the Moon was inherited from the Earth in a synestia-like scenario (Lock et al., 2018). It seems more likely that cometary Xe was delivered to the terrestrial planets after the Moon-forming event. This is because both the lunar Xe shows evidence for it, and unless the target proto-Earth was dry at the time, the giant impact that formed the Moon is expected to have stripped Earth of much of its primary atmosphere, including the noble gases (Genda & Abe 2005; Schlichting & Mukhopadhyay 2018). We view these combined lines of evidence as a means to narrow the time interval during which the giant planets migrated to soon after Moon formation, at which time the dynamical consequences of migration had all but disappeared by 4450 Ma.

Further constraints arise from lunar origin models and dynamical history of binary objects in the outer solar system. The time for the proposed giant impact that formed the Moon is contested, but may be close to 4510 Ma (Barboni et al. 2017; cf. Borg et al 2011). Our dynamical tests show that the cometary bombardment should have endured for about 30 Myr. Such an intense but short-lived flux of comets that ended by 4450 Ma would be incapable of further wholesale crustal destruction and concomitant resetting of highly retentive ages in any sampled solar system object, including lunar rocks (Nemchin et al 2009; Norman & Nemchin 2014; Hopkins & Mojzsis, 2015; Kelly et al. 2018) and vestoid meteorites (Hopkins et al 2015).

The ostensible timing of coincident impacts of large "Late Veneer"-scale singular bodies to the Earth (~3000 km diameter) (Brasser et al. 2016; Day et al. 2016) and Mars (~1000 km diameter) (Brasser & Mojzsis 2017; Bouvier et al. 2018) also conveniently explains the coterminous ages of the silicate crusts of these worlds by about 4480 Ma (Zhang 2002), but evidently no earlier. The low probability of two colossal impacts occurring nearly simultaneously on both Earth and Mars (Brasser et al. 2016) provides circumstantial evidence in support of giant planet migration as a trigger at that time. Finally, giant planet migration must have happened within the first 100 Myr of solar system history to account for the survival of Jupiter Trojan binary asteroids (Nesvorný et al. 2018).

*7.2 Implications for the emergence of life on Hadean Earth*

Our reconciled timeline for late accretion to the inner solar system now leads us to explore arguments for and against the existence of a persistent biosphere on the early Hadean Earth (Mojzsis et al. 1996; Ryder 2002). Building on related works (Maher & Stevenson 1988; Abramov & Mojzsis 2009; Abramov et al. 2013; Marchi et al. 2014) we show that annihilation of the terrestrial surface zone (including wholesale melting of the lithosphere) before ~4450 Ma was such that no stable niches for the prebiotic chemistry leading to life were possible in the first ~100 Myr. If the last global-scale crustal melting on Earth corresponds to a colossal impact at ca. 4450 (minimum mass required is $9.31 \times 10^{21}$ kg) such an event effectively sterilized the planet by eroding the hydrosphere, melting crust and creating shallow magma oceans (e.g. Schlichting & Mukhopadhyay 2018). To further elaborate on how late accretion affects the nascent surface biosphere, we employ a suite of revised global 3-D analytical codes to analyze the global thermal fields of impacts from our new late accretion mass-production functions in the time frame 4480-3500 Ma (**Figure 7**).

Results in all simulation cases show that an abating impact flux from late accretion is inadequate to sterilize the surface zone - defined here as ≥130°C for the volume of the upper 4 km of the crust - after about 4400 Ma. This conservative estimate has important implications for the origin of life on Earth. If the shift from the non-living to living world included an RNA biome as a transitional form inhabiting relatively mild aqueous surface regions (Mojzsis et al. 2001; Powner et al. 2009; Benner et al. 2012; Becker et al. 2018), such fragile proto-organisms (Bernhardt 2012) were eminently susceptible to extinction from successive regional thermal stresses to the hydrosphere from the largest impacts. In light of this, we speculate that the RNA World's top-down successors – those thermally more robust DNA-peptide microbial immigrants adapted to a colonize deep and hot crustal settings near hydrothermal vents well away from the mass- and UV-bombarded surface zone – endured the waning stages of bombardment to repopulate the planet from the bottom-up.

## 8. SUMMARY

The giant planets radially migrated after dissipation of the protoplanetary nebula around the young Sun, yet the timing of this migration is unconstrained. Here we show that radiometric ages from various meteorite classes can be reconciled with giant planet migration if it occurred before 4480 Ma. Our result resolves crater chronologies with late accretion timescales and dynamic models. It also implies that conditions on the Hadean Earth during late accretion do not preclude the emergence of an enduring biosphere as early as ~170 million years after solar system formation.

**Acknowledgements**

This work was funded by the Collaborative for Research in Origins (CRiO), which is supported by The John Templeton Foundation (principal investigator: Steven

Benner/FfAME): the opinions expressed in this publication are those of the authors, and do not necessarily reflect the views of the John Templeton Foundation. S.J.M. and N.M.K. were supported by the NASA Cosmochemistry program (Grant # NNX14AG31G) in part of this work. S.J.M. also thanks the Earth-Life Science Institute (ELSI) at the Tokyo Institute of Technology, and the Geophysical Fluid Dynamics Group at the Swiss Federal Institute of Technology (ETH) in Zürich, for their generosity in hosting him during significant phases of this research. R.B. is grateful for financial support from JSPS Kakenhi (17KK0089). O.A. and S.J.M. thank the NASA Solar System Workings program (Grant # NNH16ZDA001N). S.C.W. appreciates the support by the Research Council of Norway via the Centre of Excellence grant to the Centre of Earth Evolution and Dynamics (CEED 223272). We thank Steven Benner, Thomas Carell, Nicolas Dauphas and Michael Russell for valuable discussions that helped improve the manuscript.

**Author Information**

Correspondence and request for materials should be addressed to S.J.M. (mojzsis@colorado.edu)

**Contributions**

S.J.M. formulated the study, supervised all work, interpreted data and wrote the manuscript. R.B. developed dynamical models of accretion, helped design the study, analyzed and interpreted the results from the dynamical models and co-wrote the manuscript. N.M.K. compiled the thermochronologies of solar system objects. O.A. wrote the revised analytical thermal bombardment codes and ran simulations of the crustal evolution of the Hadean Earth using our new impact bombardment production functions. S.C.W. performed supportive crater statistical analyses and shared her crater chronology data.

The authors declare no conflict of interest.